\documentstyle[12pt]{article}

\def\a{\alpha}

\def\l{\lambda}

\def\F{\Phi}

\def\cu{{\cal U}}

\def\dag{\dagger}
\def\pro{\propto}
\def\la{\left}
\def\ra{\right}
\def\pa{\partial}

\def\Tilde#1{\widetilde {#1}}
\def\Hat#1{\rlap{\kern.10em$\widehat{\phantom G}$}#1}
\def\HAt#1{\rlap{\kern.05em$\widehat{\phantom G}$}#1}

\def\cap#1{\rlap{\kern.1em$\widehat{\phantom{G\vrule height.8em}}$}#1{}}
\def\Cap#1{\rlap{\kern.05em$\widehat{\phantom{G\vrule height.8em}}$}#1{}}

\def\ev#1{\left\langle #1\right\rangle}

\catcode`@=11
\def\underline#1{\relax\ifmmode\@@underline#1\else
        $\@@underline{\hbox{#1}}$\relax\fi}
\catcode`@=12

\def\PRL{Phys. Rev. Lett.\ }

\def\PLB{Phys. Lett. B\ }

\def\NPB{Nucl. Phys. B\ }

\def\JMP{J. Math. Phys.\ }

\def\ba{\begin{array}}
\def\ea{\end{array}}
\def\be{\begin{equation}}
\def\ee{\end{equation}}
\def\bdm{\begin{displaymath}}
\def\edm{\end{displaymath}}
\def\bea{\begin{eqnarray}}
\def\eea{\end{eqnarray}}
\def\nl{\nonumber \\}

\def\by{\over}
\def\lb{\label}
\def\bl#1{(\ref{#1})}
\def\sp{~~~~~}

\def\vsp{\\ \vglue 0.2in}
\def\au{{\it Department of Physics\\
          University of Arizona, Tucson, AZ 85721}}

\def\cu{{\it Department of Physics\\
          University of Colorado, Boulder, CO 80309}}

\def\bt#1#2#3#4#5
        {\begin{titlepage}
         \centerline{\hfill {#1}}
         \centerline{\hfill {#2}}
         \centerline{\hfill {#3}}
         \begin{center}\vskip .2in
         {\large\bf {#4}}\\[.3in]
         {\bf {#5}}\\[.3in]
         {\bf ABSTRACT}\\
         \end{center}
         \begin{quotation}}
\def\et
         {\end{quotation}
          \end{titlepage}}

\newcounter{sxn}

\newcounter{axn}

\def\br{\begin{thebibliography}{abc}}
\def\er{\end{thebibliography}}
\def\rf{\bibitem}
\def\fr#1{\cite{#1}}
\begin{document}

\bt{AZPH-TH/93-01, COLO-HEP/303}{hep-th/9212158}{January 1993}
{Difficulties in Inducing a Gauge Theory at Large $N$}{B. S.
Balakrishna \vsp \footnote{Address for correspondence. Email:
bala@haggis.colorado.edu}\cu \vsp and \vsp \au}

It is argued that the recently proposed Kazakov-Migdal model of
induced gauge theory, at large $N$, involves only the zero area Wilson
loops that are effectively trees in the gauge action induced by the
scalars. This retains only a constant part of the gauge action
excluding plaquettes or anything like them and the gauge variables
drop out.

\et

Recently, Kazakov and Migdal made an attempt to induce large $N$ QCD
from a lattice model\fr{qcd}. This model is far simpler than the usual
lattice QCD as it has no kinetic term for the gauge variables.  The
hope was that the scalars present in the model may somehow generate
this kinetic term and lead to a nontrivial gauge theory. The
simplicity of the model let the authors of \fr{qcd} to solve the model
exactly at large $N$. This exact solution was expected to correspond
to large $N$ QCD near some critical point. Initial attempts involving
a quadratic potential for the scalar variables did not
succeed\fr{gro}, but the hope remained that a nontrivial potential may
perhaps induce large $N$ QCD. Numerical\fr{num} and analytical\fr{ana}
means suggested that there do exist critical points in the model
exhibiting nontrivial scaling behavior.

A significant amount of work\fr{lit} has already been done in this
regard, but it is still not clear what the large $N$ result really
corresponds to. Is it a theory of scalars interacting with the gauge
fields? If so, can the scalars be made heavy enough to decouple from
the gauge variables to induce QCD? The effective action obtained by
integrating away the scalars should hold a clue to these questions.
It is an action for the gauge variables involving their gauge
invariant combinations, the Wilson loops, of all sorts. But, there is
a danger that the nontrivial loops of interest might get suppressed
for large $N$, leaving only those bounding zero area. Because this
excludes plaquettes or anything like them, the gauge variables
decouple. It would still be a highly nontrivial matter to solve the
model at large $N$. Here, we present arguments that support this
unfortunate scenario.

The Kazakov-Migdal model is defined by the following action defined on
a $D$ dimensional hypercubic lattice:
\be
S ~=~ N^2\sum_x{\rm tr}V(\F(x)) -
N^2\sum_{\ev{xy}}{\rm tr}\la[\F(x)U(xy)\F(y)U^\dag(xy)\ra].   \lb{s}
\ee
The lattice sites are labeled by $x$ and the links by $\ev{xy}$. Gauge
variables are represented by $N\times N$ unitary matrices, $U(xy)$'s,
associated with the links. For large $N$, this could be regarded as a
model based on U$(N)$ rather than SU$(N)$. $\F$ is an $N\times N$
hermitian matrix representing a scalar in the adjoint representation
that could have a nonzero trace. One factor of $1/N$ is implicit in
the definition of trace. Integration over $U$ can be performed
separately on every link. The integral over the link $\ev{xy}$ depends
only on the eigenvalues of $\F(x)$ and $\F(y)$. Here it is easier to
work with the traces, ${\rm tr}\F^i, i=1,2,\cdots$, rather than the
eigenvalues. Only $N$ of them are independent, but it is convenient to
keep all of them. The effective action for the scalars arising from
the $U-$integrations is of the form
\be
S_{\rm eff} ~=~ N^2\sum_x{\rm tr}V(\F(x)) -
N^2\sum_{\ev{xy}}J(\F(x),\F(y)),                         \lb{s2}
\ee
where $J$ comes from the link integral,
\be
{\rm exp}\la[N^2J(\F(x),\F(y))\ra] ~=~ \int dU~{\rm exp}\la[N^2{\rm
tr}\la(\F(x)U\F(y)U^\dag\ra)\ra].                        \lb{j}
\ee
An explicit expression for $J$ is not needed for our purpose. Note
that $J$ can be regarded as an independent function of all the traces
of $\F(x)$'s and $\F(y)$'s. This follows from expanding the
exponential inside the integral and integrating over $U$\fr{crz}. An
expansion of $J$ in its arguments is of the form
\be
J(\F(x),\F(y)) ~=~ \sum_{n\a}J_{n\a}(\F(x),\F(y)),
\ee
where $J_{n\a}$ is a product of the various traces that together
involve $n$ of the $\F(x)$'s and $n$ of the $\F(y)$'s, $\a$ labeling
the different possibilities. For instance, at the $x$ end or the $y$
end,
\be
J_{1\a} ~\pro~ {\rm tr}\F, \sp J_{2\a} ~\pro~ {\rm tr}\F^2,~ ({\rm
tr}\F)^2, \sp J_{3\a}(\F) ~\pro~ {\rm tr}\F^3,~ {\rm tr}\F^2{\rm
tr}\F,~ ({\rm tr}\F)^3.
\ee
For large $N$, $J$ has no explicit dependence on $N$ except for that
absorbed into the definition of trace. Expansion of the partition
function by perturbing $J$ generates various graphs, trees as well as
loops. For every choice of $n$ and $\a$, $J$ provides us with an edge
for constructing the graphs. Graphs arise when these edges are joined
together by contracting the $\F$'s. As it turns out, only the trees
contribute to the large $N$ limit.

Let us first discuss a primary edge of order $n=2$ having ${\rm
tr}\F^2$ attached to the ends. To make the counting of $N$'s easier,
let us scale $\F$ as $\F\to\F/N$ leading to a factor $1/N^2$ in front
of this edge term. Each such edge now supplies $1/N^2$. A vertex of
order $p$, formed when $p$ edges are glued at a site by contracting
the $\F$'s, involves a summation over the $N^2$ components of $\F$ and
hence contributes a factor $N^2$. This is true for the open ends
($p=1$) and the joints ($p=2$) as well. The $\F$ propagator involved
in the $\F$ contractions has one factor of $N$ after scaling that
compensates the $1/N$ factor implicit in the definition of trace in
${\rm tr}\F^2$. If the coupling constants of the potential also take
part in the gluing process, an exercise in the zero dimensional matrix
field theory of $\F$ defined at a site shows that $N^2$ is the leading
contribution from a vertex. Note that the coupling constants could
carry $1/N$'s after scaling. The product of $N^2$'s is thus
$(N^2)^{V-E}$ at most, where $V$ is the total number of vertices and
$E$ is the total number of edges. This equals $(N^2)^{1-L}$ for a
connected graph having $L$ loops. Clearly, only the trees are
responsible for the leading, order $N^2$, behavior. The potential is
not required to be even in $\F$ for this discussion. $\F$ may have a
nonzero expectation value that is proportional to the identity with a
factor $N$ due to scaling. These conclusions hold good for all the
primary edges, that is, edges of order $n$ having one trace of the
form $N^{2-n}{\rm tr}\F^n$ ($N$'s coming from scaling) attached to the
ends.

The remaining edges have more than one traces attached to the ends.
After scaling, an order $n$ edge receives a factor $1/N^2$ for itself
and a factor $N^{2-n}$ at the ends that carry the traces. One may view
this as a primary edge having some more traces attached. If the
additional traces that are of the form $N^{-m}{\rm tr}\F^m$ are
self-contracted, they simply contribute a factor unity. Then the
arguments are the same as that for the primary edges and only trees
contribute at large $N$. In general, a whole branch of a tree can grow
from the additional traces and hence from all the traces. The branch
attached to a trace $N^{2-m}{\rm tr}\F^m$ contributes at most $N^2$
like an open end of a primary edge. This suggests that ${\rm tr}\F^m$
supplies a factor $N^m$ and hence the product of traces a factor
$N^n$. This along with $N^{2-n}$ coming from scaling combines to give
$N^2$. Each end of an edge thus carries a $N^2$ and $1/N^2$ is
supplied by the edge itself, giving together $N^2$, the leading order
for the whole tree. This argument is applicable to the branches that
grow from the traces as well. Any other way of constructing the graphs
is found to be of higher order in $1/N$. Thus, in general, only trees
contribute at large $N$.

There is one more possible source of $N$. This is due to the presence
of $N$ independent traces. But as we have seen, all the traces are on
a similar footing in defining the edges and generating graphs, and it
is unlikely for $N$ of them to enhance the loops. This also follows
from the fact that the sum of all the trees constructed above gives a
result that agrees with that of the saddle point method. To see this,
let us derive an expression that sums up the leading trees. In the
process, we obtain an equation that is analog of the saddle point
equation of the lattice model. First, introduce a $\l$ for each of the
traces, that is $\l_i$ for ${\rm tr}\F^i$. As noted earlier, it is
possible to regard $J$ as an independent function of all the traces.
Expanding $J$ around $\l$ gives
\be
J\la(\F(x),\F(y)\ra) ~=~ J(\l) + \sum_i\la({\rm tr}\F^i(x) + {\rm
tr}\F^i(y) - 2\l_i\ra)\pa_iJ(\l)/2 + \Tilde{J},
\ee
where $\pa_iJ(\l)=\pa J(\l)/\pa\l_i$ and $\Tilde{J}$ is the remainder.
For the trees constructed earlier, the open ends carry a weight given
by the product of the expectation values (in the zero dimensional
matrix field theory at a site) of all the traces attached. Let us
choose $\l_i$ to be the expectation value of ${\rm tr}\F^i$ when the
above expansion of $J$ is used in Eq. \bl{s2} without $\Tilde{J}$.
Including $\Tilde{J}$ as a perturbation, note that the graphs
generated exclude the trees constructed earlier. This is because
$\Tilde{J}$ depends on the traces as ${\rm tr}\F^i-\l_i$ and our
choice of $\l$ makes it not possible for ${\rm tr}\F^i-\l_i$ to create
open ends of leading order. This suggests that the lowest order (that
is, without $\Tilde{J}$) should be summing up the trees. This sum is
\be
W\la(D\pa J(\l)\ra) + DJ(\l) - D\sum_i\l_i\pa_iJ(\l)        \lb{w}
\ee
per site, where $W$ is defined by
\be
{\rm exp}\la[N^2W(j)\ra] ~=~ \int d\F~{\rm exp}\la[-N^2{\rm tr}V(\F) +
N^2\sum_ij_i{\rm tr}\F^i\ra].
\ee
Derivative of $W(j+D\pa J(\l))-\sum j\l$ with respect to $j$ at $j=0$
gives the leading weight associated with the open ends. This should
vanish by our choice of $\l$, giving
\be
\pa_iW\la(D\pa J(\l)\ra) - \l_i ~=~ 0,
\ee
where $\pa_iW(j)=\pa W(j)/\pa j_i$. This equation determining $\l$ is
analogous to the saddle point equation of the lattice model and is
equivalent to the latter at large $N$. $W$ at large $N$, when used in
\bl{w}, gives the free energy per site at the saddle point. The integral
determining $W$ is dominated at a saddle point for large $N$ leading
to a saddle point equation of its own, which agrees with that of the
lattice model when $j=D\pa J(\l)$. The equation derived above simply
tells us to identify $j$ with $D\pa J(\l)$. It also extremizes \bl{w}.
Second and higher derivatives of $W$ evaluated at $D\pa J(\l)$ help us
sum the branches that could grow from the vertices of loops.

These conclusions reinforce ones conviction that the saddle point
method is closely related to summing the tree graphs. This can be
translated to the statement that, at large $N$, only trivial Wilson
loops bounding zero area contribute to the gauge action obtained by
integrating away the scalars. The analysis so far involved the
effective scalar action. The gauge action can be studied by perturbing
the link term of \bl{s}. The link terms when joined by contracting the
$\F$'s also generate graphs. To relate the two types of graphs, expand
the exponential inside the $U-$integral in Eq. \bl{j} as well as the
exponential of $N^2J$. This leads to a series of relations of the
kind,
\bea
N^2\int dU~{\rm tr}\la(\F(x)U\F(y)U^\dag\ra) &=& N^2J_1, \nl
{1\by 2}N^4\int dU~\la[{\rm tr}\la(\F(x)U\F(y)U^\dag\ra)\ra]^2 &=&
N^2J_2 + {1\by 2}N^4J_1^2,
\eea
where $J_n=\sum_{\a}J_{n\a}$. One may view these as decomposing the
$U-$integrals into connected pieces\fr{crz}. Nontrivial Wilson loops
do not contribute at large $N$ simply because there are no such $\F$
contractions to do the job. The first relation lets us replace a $J_1$
edge, when present alone on a link, by $\int dU~{\rm
tr}(\F(x)U\F(y)U^\dag)$. Trees made of $J_1$ edges (one per link) are
thus the trivial Wilson loops bounding zero area. To understand the
other relations, it is convenient label the $\F$'s such that those
connected by $U$ have the same label. This defines a labeling of the
$\F$'s for the $J$'s. For instance, a labeling in $[{\rm
tr}(\F(x)U\F(y)U^\dag)]^2$ relates to two $J_1$ and one $J_2$ edges.
Integrating over $U$ and collecting the connected pieces, one finds
that a $J$ edge carries the same set of labels at its $x$ and $y$
ends. A nontrivial Wilson loop defines a sequence of labels along the
loop corresponding to a sequence of $\F$ contractions. Such a sequence
can not arise from the trees made of $J$ edges. It is thus not
possible to feel the presence of nontrivial Wilson loops at large $N$.
For the trivial ones, the product of $U$'s multiplies to unity and the
gauge variables decouple. To see the effects of plaquette like terms,
one needs to go to subleading orders. This is true for a fundamental
scalar as well and perhaps is also true for other models where the
angular variables decouple from the effective scalar action.  A model
with $N$ fundamental scalars is free of these problems, but it is very
difficult to solve.

The large $N$ result could still be useful for discovering a double
scaling limit which is quite a generic phenomenon of scalar field
theories\fr{bal}. The present analysis helps us in identifying the
$1/N$ corrections. The tree graphs themselves receive corrections at
the vertices, the loop graphs become significant and the link integral
gets modified. $1/N$ expansion in these models is possibly related to
a strong coupling expansion of the induced gauge theory. The large $N$
result is just a constant part of the induced gauge action, but it
does have implications for the subleading terms. The tree graphs
modify the vertices of all the loops and the gauge coupling will be
strongly dependent on them. The lattice model involving a fundamental
scalar will provide a simpler framework to study these matters.

To summarize, the Kazakov-Migdal model at large $N$ retains only a
constant part of the effective action for the gauge variables obtained
by integrating the scalars. This arises from the zero area Wilson
loops that are effectively trees. To feel the presence of plaquette
like terms, one needs to go beyond the leading order. The large $N$
result could still be useful as it modifies indirectly the strength of
the subleading terms and is expected to contribute to the gauge
coupling constant as well as critical behavior.

This work is supported by NSF Grant PHY-9023257.

\br
\rf{qcd}
V. A. Kazakov and A. A. Migdal, preprint PUPT-1322, LPTENS-92/15, to
be published in \NPB.
\rf{gro}
D. Gross, \PLB 293, 181 (1992).
\rf{num}
A. Gocksch and Y. Shen, \PRL 69, 2747 (1992).
\rf{ana}
A. A. Migdal, preprint PUPT-1323.
\rf{lit}
I. I. Kogan, G. W. Semenoff and N. Weiss, \PRL 69, 3435 (1992); A. A.
Migdal, preprint PUPT-1332; M. Caselle, A. D'Adda and S. Panzeri, \PLB
293, 161 (1992); I. I. Kogan, A. Morozov, G. W. Semenoff and N. Weiss,
preprint UBCTP 92-26, ITEP M6/92; S. Khokhlachev and Yu. Makeenko,
preprint ITEP-YM-5-92; A. A. Migdal, preprint LPTENS-92/22; I. I.
Kogan, A.  Morozov, G. W. Semenoff and N. Weiss, preprint UBCTP 92-27,
ITEP M7/92; M.I. Dobroliubov, I.I. Kogan, G.W. Semenoff and N. Weiss,
preprint PUPT 1358, UBCTP92-032.
\rf{crz}
Integration over $U$'s and $U^\dag$'s is given by a sum of products of
Kronecker delta symbols. For details see, M. Creutz, \JMP 19, 2043
(1978).
\rf{bal}
B. S. Balakrishna, preprint AZPH-TH/92-43, COLO-HEP/295.
\er

\end{document}